\documentclass{IEEEcsmag}

\usepackage[colorlinks,urlcolor=blue,linkcolor=blue,citecolor=blue]{hyperref}
\expandafter\def\expandafter\UrlBreaks\expandafter{\UrlBreaks\do\/\do\*\do\-\do\~\do\'\do\"\do\-}
\usepackage{upmath,color}
\usepackage{xurl}
\usepackage{xcolor}
\jvol{jvol}
\jnum{jnum}
\paper{paper}
\jmonth{Month}
\jname{Publication Name}
\jtitle{title}
\pubyear{2023}

\usepackage{datetime2}
\usepackage{doi}

\usepackage{datetime}

\newenvironment{titemize} 
        {\begin{list}{\labelitemi}{
                \setlength{\topsep}{1pt}
                \setlength{\parskip}{2pt}
                \setlength{\itemsep}{2pt}
                \setlength{\parsep}{0pt}
                \setlength{\leftmargin}{10pt}
                \setlength{\labelwidth}{10pt}
        }}
        {\end{list}}

\definecolor{linkblue}{RGB}{0,0,180}
\hypersetup{
     colorlinks=true,
     citecolor=linkblue,
     filecolor=black,
     linkcolor=linkblue,
     urlcolor=linkblue
}

\begin{document}

\title{A cast of thousands: How the IDEAS Productivity project has advanced software productivity and sustainability}

\markboth{THEME/FEATURE/DEPARTMENT}{THEME/FEATURE/DEPARTMENT}

\author{%
  \IEEEauthorblockN{%
    {Lois Curfman McInnes}\IEEEauthorrefmark{1},
    Michael Heroux\IEEEauthorrefmark{2},
    David E.~Bernholdt\IEEEauthorrefmark{3},
    Anshu Dubey\IEEEauthorrefmark{1},
    Elsa Gonsiorowski\IEEEauthorrefmark{4},
    Rinku Gupta\IEEEauthorrefmark{1},
    Osni Marques\IEEEauthorrefmark{5},
    J.~David Moulton\IEEEauthorrefmark{6},
    Hai Ah Nam\IEEEauthorrefmark{5},
    Boyana Norris\IEEEauthorrefmark{7},
    Elaine M. Raybourn\IEEEauthorrefmark{2},
    Jim Willenbring\IEEEauthorrefmark{2},
    Ann Almgren\IEEEauthorrefmark{5},
    Ross Bartlett\IEEEauthorrefmark{2},
    Kita Cranfill\IEEEauthorrefmark{3},
    Stephen Fickas\IEEEauthorrefmark{7},
    Don Frederick\IEEEauthorrefmark{4},
    William Godoy\IEEEauthorrefmark{3},
    Patricia Grubel\IEEEauthorrefmark{6},
    Rebecca Hartman-Baker\IEEEauthorrefmark{5},
    Axel Huebl\IEEEauthorrefmark{5},
    Rose Lynch\IEEEauthorrefmark{1},
    Addi Malviya Thakur\IEEEauthorrefmark{3},
    Reed Milewicz\IEEEauthorrefmark{2},
    Mark C.~Miller\IEEEauthorrefmark{4},
    Miranda Mundt\IEEEauthorrefmark{2},
    Erik Palmer\IEEEauthorrefmark{5},
    Suzanne Parete-Koon\IEEEauthorrefmark{3},
    Megan Phinney\IEEEauthorrefmark{6},
    Katherine Riley\IEEEauthorrefmark{1},
    David M. Rogers\IEEEauthorrefmark{3},
    Ben Sims\IEEEauthorrefmark{6},
    Deborah Stevens\IEEEauthorrefmark{1},
    Gregory R. Watson\IEEEauthorrefmark{3}
}
\IEEEauthorblockA{\IEEEauthorrefmark{1}Argonne National Laboratory, Lemont, Illinois, 60439, USA}
\IEEEauthorblockA{\IEEEauthorrefmark{2}Sandia National Laboratories, Albuquerque, New Mexico 87185, USA}
\IEEEauthorblockA{\IEEEauthorrefmark{3}Oak Ridge National Laboratory, Oak Ridge, Tennessee 37831, USA}
\IEEEauthorblockA{\IEEEauthorrefmark{4}Lawrence Livermore National Laboratory, Livermore, California 94550, USA}
\IEEEauthorblockA{\IEEEauthorrefmark{5}Lawrence Berkeley National Laboratory, Berkeley, California 94720, USA}
\IEEEauthorblockA{\IEEEauthorrefmark{6}Los Alamos National Laboratory, Los Alamos, New Mexico 87545, USA}
\IEEEauthorblockA{\IEEEauthorrefmark{7}University of Oregon, Portland, Oregon 97403, USA}
}

\begin{abstract}
Computational and data-enabled science and engineering are revolutionizing advances throughout science and society, at all scales of computing.  For example, teams in the U.S. DOE Exascale Computing Project have been tackling new frontiers in modeling, simulation, and analysis by exploiting unprecedented exascale computing capabilities—building an advanced software ecosystem that supports next-generation applications and addresses disruptive changes in computer architectures.  However, concerns are growing about the productivity of the developers of scientific software, its sustainability, and the trustworthiness of the results that it produces.  Members of the IDEAS project serve as catalysts to address these challenges through fostering software communities, incubating and curating methodologies and resources, and disseminating knowledge to advance developer productivity and software sustainability.
This paper discusses how these synergistic activities are advancing scientific discovery—mitigating technical risks by building a firmer foundation for reproducible, sustainable science at all scales of computing, from laptops to clusters to exascale and beyond.
\end{abstract}

\maketitle
\section{Introduction}

Mathematics-based advanced computing has become ubiquitous across scientific domains, and software is its backbone. 
In addition, software is essential to the operation of experiments and scientific user facilities and to many 
discoveries happening every day at these facilities. 
The sustainability of this tremendous investment in
software is as important as that of other kinds of scientific instruments and deserves the same sort of attention. 
Moreover, next-generation scientific challenges---coupling simulations with data from experiments and observations via {\em integrated research infrastructure},\footnote{Emerging workloads require 
integration among computing, experimental and observational facilities, or {\em integrated research infrastructure}, \url{https://doi.org/10.2172/1863562}.}
exploiting advanced hybrid computing architectures via new algorithms for high-performance computing (HPC),  bridging scales and domains, and enhancing insight through emerging AI/ML tools---have brought concerns about software quality, developer productivity, scientific reproducibility, and software stewardship into sharp focus.\footnote{The ASCR Workshop on the Science of Scientific Software Development and Use considered challenges and opportunities for next-generation scientific software, {\doi{10.2172/1846009}}.
}

In response to these challenges, 
grass-roots groups are fostering communities of practice, 
where like-minded people share information and experiences on effective approaches for creating, sustaining, and collaborating via scientific research software. These groups articulate key issues to stakeholders, agencies, and the broader community to effect changes in policies, funding, and reward structure, while advancing understanding of the importance of high-quality software to effective collaboration and the integrity of computational research.  
Groups such as CIG\footnote{CIG (geoscience): \url{https://geodynamics.org}} and MolSSI\footnote{MolSSI (molecular science): \url{https://molssi.org}} focus on the needs of particular application areas, while complementary organizations such as the Software Sustainability Institute,\footnote{SSI: \url{https://www.software.ac.uk}} Carpentries,\footnote{Carpentries: \url{https://carpentries.org}} Research Software Alliance,\footnote{ReSA: \url{http://www.researchsoft.org}} and the Center for Scientific Collaboration and Community Engagement\footnote{CSCCE: \url{https://www.cscce.org}} address broader concerns~\cite{community-organizations:2019}. 
Moreover, 
the growing community of {\em research software engineers (RSEs)}\footnote{\url{https://us-rse.org}, \url{https://society-rse.org}} is raising awareness of software's critical role in research and advancing advocacy and resources for those who regularly use programming expertise to advance research. 

The Interoperable Design of Extreme-scale Application Software (IDEAS) multiphase project\footnote{\url{https://ideas-productivity.org}} is sponsored by the U.S. Department of Energy (DOE) to advance scientific software productivity and  sustainability.  The IDEAS project is the first of
its kind in the United States with its entire focus on incubating, curating, and disseminating knowledge and methodologies about the sustainment of scientific software.  
The IDEAS team, in collaboration with complementary organizations, is working toward a scientific software community culture that invests in and benefits from an explicit focus on developer productivity and software sustainability, adapting and adopting best practices from the broader software community and incubating our community-specific contributions to these pursuits.

While IDEAS and other synergistic efforts have made tremendous progress, there is still a long way to go.
In order to address growing challenges in next-generation scientific computing, an explicit, concerted effort is needed more urgently than ever to incubate, curate, and disseminate innovations across groups, fostering greater adoption of effective practices, processes, and tools for scientific software development and use.  
These advances are needed to improve transparency and reproducibility, as required for open, collaborative computational and data-enabled science and engineering, including AI/ML, to address pressing problems throughout our world. 

\section{A Brief History of IDEAS}

The IDEAS project began in 2014, sponsored by the DOE Office of Science as a partnership between the Offices of Advanced Scientific Computing Research (ASCR) and Biological and Environmental Research (BER) to address challenges in multiscale, multiphysics terrestrial ecosystem modeling. We now refer to this original phase of the project as IDEAS-Classic\footnote{\url{https://ideas-productivity.org/ideas-classic}}---a dual-focused effort to create a math libraries ``software development kit,'' 
called xSDK,\footnote{\url{http://xsdk.info}} 
and to build a community committed to elevating the importance of and the abilities to realize developer productivity and software sustainability improvement.
With the inception of the U.S. Exascale Computing Project (ECP)~\cite{ecp-kothe-lee-qualters-2019},\footnote{\url{https://www.exascaleproject.org}} IDEAS became in 2017 the IDEAS-ECP project,\footnote{\url{https://ideas-productivity.org/ideas-ecp}} 
addressing productivity and sustainability efforts throughout the ECP community, and the xSDK became a separate effort under the ECP Software Technology umbrella. 
We are working toward the project's post-ECP existence again as IDEAS, a distributed but collective effort across multiple DOE-sponsored projects.
Meanwhile, in 2019 the BER-funded IDEAS-Watersheds project\footnote{\url{https://ideas-productivity.org/ideas-watersheds}} grew out of IDEAS-Classic, with emphasis on accelerating watershed science through a community-driven software ecosystem.
Throughout this report, we use the term {\em IDEAS} by itself 
to refer to overarching IDEAS initiatives 
and {\em IDEAS-ECP} when 
referring to IDEAS activities that were conducted as part of the now-completed ECP project.

{\bf Challenges in developer productivity and software sustainability in large-scale team science.} To meet the needs of next-generation science, ECP brought together the developers of 70 distinct software technologies\footnote{The 2022 ECP Software Technology Capability Assessment Report, \doi{10.2172/1888898}, described ECP software capabilities and activities.} to create a rich and sustainable scientific software ecosystem~\cite{SWEcosystems:NCS2021} that supported a diverse suite of applications in chemistry, materials, energy, Earth and space science, data analytics, optimization, AI, and national security. 
Projects targeted portable performance across multiple exascale computer architectures (capable of exceeding a quintillion, or $10^{18}$, calculations per second).
ECP's aggressive goals required intensive software refactoring and development, involving more than 1,000 researchers across DOE labs and collaborating universities, as well as partnerships with DOE computing facilities, industry, and other US agencies. The IDEAS-ECP project was tasked with helping  mitigate the challenges of software development in this environment\footnote{A 2020 IDEAS report, \doi{10.2172/1606662}, enumerated technical and cultural challenges in scientific software productivity and sustainability.} and thereby enhance the productivity and sustainability of the software ecosystem.

ECP had an ambitious schedule, where teams were asked to deliver performant and portable software for newly developed hardware, which in turn required improving software practices. For many teams, this marked an additional focus on long-term software productivity and sustainability in addition to short-term scientific output.\footnote{A 2020 paper provided insights into software practices of a variety of HPC projects, \doi{10.1177/1094342019899451}.}

ECP showed that it is possible to build productive, cross-disciplinary teams that deliver combined advances in the adoption of new (and disruptive) hybrid computing hardware,\footnote{The DOE machines Frontier (\url{https://www.olcf.ornl.gov/frontier}), Aurora (\url{https://www.alcf.anl.gov/aurora}, and El Capitan (\url{https://computing.llnl.gov/about/newsroom/road-el-capitan-4}) employ GPU-accelerated architectures to enable transformative advances in science.} programming models, mathematical libraries, data management and visualization packages, development tools, and application-specific functionality to achieve unprecedented capabilities in modeling, simulation, and analysis. However, complexity grows with project size and scope---measured by user/developer community size, project team interactions, and lines of code.
Bringing together the broad range of expertise needed to achieve ECP objectives also revealed new challenges: 
aligning the goals of cross-disciplinary project teams, where contributors are motivated by a variety of diverse priorities and have difficulty in capturing different milestones and objectives; 
fostering more effective communication across teams, where members may differ on how to best accomplish shared goals; and scaling the {\em diffusion of  innovations}~\cite{Diffusion-of-Innovations-Rogers} across projects. Diffusion of innovations is concerned with the spread or propagation of ideas and new technologies. We have applied this theoretical perspective to devise a multi-pronged strategy and to actively engage with early adopters of productivity methods and seek champions within the community who are willing to advocate for a culture of productivity. 
We help catalyze early adopters and ensure that best practices continue to grow through the other phases of the diffusion process. These insights have been continually shaping the evolution of our priorities.

{\bf IDEAS strategy: Reducing technical risks through culture change.} 
Aligned with the need for longer-term perspectives, 
members of the multi-institutional IDEAS team, including staff embedded in DOE research teams and advanced computing facilities, many of whom are RSEs~\cite{MundtEtAl2022}, are working to qualitatively change the software development culture of computational science---improving the practice of and quality of scientific software. We are pursuing a multipronged strategy: 
\begin{titemize}
\item fostering software communities,
\item incubating and curating methodologies and resources, and
\item disseminating knowledge to advance developer productivity and software sustainability.
\end{titemize}

\begin{figure*}[tbh]
\vspace{-0.1in}
  \begin{centering}
\includegraphics[width=0.90\textwidth]{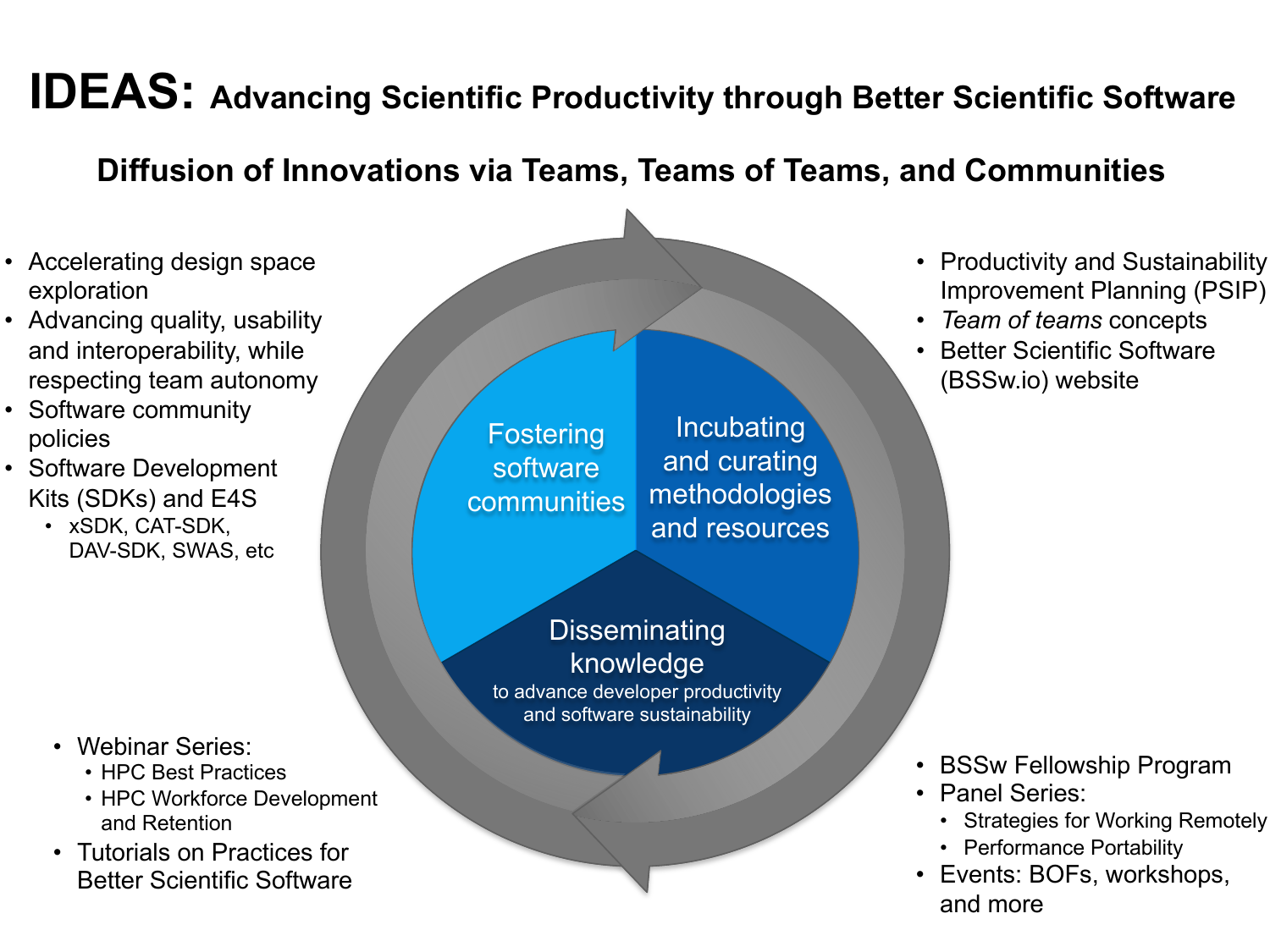}
\vspace{-0.1in}
  \caption{The IDEAS team engages with DOE application and software teams---and the broader HPC community---to reduce technical risks and build a firmer foundation for next-generation computational science. Our strategy of being a persistent catalyst means we must always incubate, curate, disseminate, and repeat---ensuring we transfer our discoveries and knowledge to the broader community to create sustainable and evolving impact.}
  \label{fig:ideas-bigpicture}
  \end{centering}
\end{figure*}

Figure~\ref{fig:ideas-bigpicture} illustrates an overview of IDEAS activities and interactions with the computational science community, 
where the IDEAS project is a focal point for resources on scientific software practices. While working closely with a variety of teams that need direct engagement, we identify and promote to others better practices that we observe with one team, increasing the rate of improvement for all. 
Our focus on best practices in scientific software, as well as fostering (in partnership with the DOE community) the development and use of software ecosystems to generate trusted computational results, enables scientists to engage effectively in their areas of expertise. At the same time, because many people beyond ECP and DOE face similar software challenges and because project participants pursue ongoing work with diverse groups, our approach also serves the broader HPC community. 

\section{Fostering software communities}
Since its inception, the IDEAS project has prioritized bringing together software teams to share knowledge and coordinate activities in a way that encourages communities to self-organize.
This work has been particularly important for teams who are developing software ecosystems, where interoperability and compatibility are essential, yet 
it is not traditionally common for developers and users of related technologies to collaborate closely across independent projects at the community level.  
While the IDEAS-Classic project founded the original software development kit (the math libraries SDK called xSDK\footnote{The xSDK brings together the developers of various math libraries to improve code quality and compatibility, \doi{https://doi.org/10.14529/jsfi170104}.}),
the primary role of IDEAS in community incubation has been in promoting the SDK approach---including methodologies to improve developer productivity and software quality as needed for robust software ecosystems---to other software communities, often using the SDK-Tools resources initiated during the IDEAS-ECP project.\footnote{\url{https://betterscientificsoftware.github.io/SDK-Tools}}

From the beginning with the xSDK, we have found that activities conducted within an intentional product community (of the developers of related software packages) have been effective at accelerating design space exploration and making compatible collections of libraries and tools that benefit users, computing facilities, and the product development teams themselves. Similarly, we have found that community software policies\footnote{\url{https://e4s-project.github.io/policies.html}}~\footnote{\url{https://xsdk.info/policies}} are important to advancing the quality, usability, and interoperability of related software technologies, while still supporting the autonomy of diverse teams that naturally have different drivers and constraints. 
For example, software advances that were motivated by xSDK community policies\footnote{ \url{https://bssw.io/blog_posts/building-community-through-software-policies}} (on installation, testing, documentation, accessibility, error handling, and more) 
overcame prior incompatibilities among a variety of high-performance numerical libraries and are improving the overall quality of the software ecosystem.  These improvements now enable the latest algorithmic advances of various math libraries to be used in combination by diverse applications, including simulations of reactive transport, radiation hydrodynamics, climate, fluid physics in wind farms, and fusion energy.\footnote{ \url{https://sinews.siam.org/Details-Page/xsdk-building-an-ecosystem-of-highly-efficient-math-libraries-for-exascale}}

Each product community has different characteristics and needs, so the activities within communities vary accordingly. For example, the math libraries community (via the xSDK) has focused substantial effort on achieving and maintaining interoperability among member products. In contrast, the data and visualization community (via the DAV-SDK) has focused on managing common versions of third-party dependencies and establishing continuous integration (CI) testing.\footnote{\url{https://bssw.io/blog_posts/bright-spots-team-experiences-implementing-continuous-integration}} 
The xSDK community thrived under ECP and continues. Other communities that benefited from observing and emulating the xSDK approach and adopting the SDK philosophy include the following.
\begin{titemize}
    \item {\bf Extreme-scale Scientific Software Stack (E4S):} As a curated stack that incorporates the various topical SDKs (including programming models and runtimes, math libraries, data and visualization libraries, and development tools), E4S\footnote{\url{http://e4s.io}}
    \footnote{\scriptsize \url{https://collegeville.github.io/CW20/WorkshopResources/WhitePapers/heroux-willenbring-shende-coti-spear-et-al-E4S.pdf}} relies on the Spack package manager.\footnote{\url{https://spack.io}} E4S facilitates the combined use of independent software packages by application teams, while also improving transparency and reproducibility of computational results. 
    \item {\bf ECP Software Development Kits (SDKs):}  The SDK approach of establishing collaborative structures for product communities grew from the original IDEAS-Classic xSDK. ECP extended the SDK approach to other areas, including programming models, development tools, data and visualization, and workflows.
    \item {\bf Code Analysis and mining Tools (CAT) SDK:} Growing from IDEAS work to understand team software practices, the CAT-SDK\footnote{\url{https://cat-sdk.github.io}} 
    is the most recent contribution to SDK community building. CAT-SDK---a suite of analysis tools (including those for analyzing repositories, source code, and pull requests)---enables software teams to gain insight into the day-to-day programming aspects of projects to help understand and improve development processes. For example, the MeerCat tool---which can check that documentation standards are being met and alleviate some of the burden of formal reviews of Git pull requests---is helping with a documentation upgrade throughout the Flash-X\footnote{\url{https://flash-x.org}} multiphysics code, which is part of the ECP ExaStar application. 
    \item {\bf The Center for Sustaining Workflows and Application Services (SWAS):} As a follow-on to the ECP ExaWorks project, SWAS\footnote{\url{https://swas.center}} brings together academia, national labs, and industry to create a sustainable software ecosystem supporting software and services used in workflows, and workflow orchestration software itself.
\end{titemize}
The impact of these multiteam efforts extends far beyond what ECP has funded directly, including teams from many organizations who have brought their products and independent funding to the activities.
For example, several math libraries external to ECP and DOE (with funding from the U.S. National Science Foundation and various European efforts) have joined the xSDK, finding benefits in community collaboration.  Likewise, E4S includes a variety of packages beyond those funded by DOE.
\section{Incubating and curating methodologies and resources}

Development and dissemination of best practices is a fundamental requirement for  improving developer productivity and software sustainability. The IDEAS team has many years of collective experience in producing and maintaining high-quality HPC scientific software. Much of this knowledge is gathered from the broader software engineering community and customized for specific needs of scientific software on topics such as agile processes, collaboration via revision control workflows, reproducibility, scientific software design, refactoring, and testing. 
A key focus is promoting continuous technology refreshment.\footnote{\scriptsize \url{https://bssw.io/blog_posts/continuous-technology-refreshment-an-introduction-using-recent-tech-refresh-experiences-on-visit}} 

{\bf Productivity and Sustainability Improvement Planning (PSIP).}  To help teams take action on improving their software practices, while respecting that they have limited time and resources and must focus on making progress on the core functionality of their software technologies and applications, we devised PSIP\footnote{\url{https://bssw.io/psip}}---a light\-weight workflow where teams incrementally and iteratively upgrade software practices~\cite{Heroux:2020:LSP}.
As a result of a multipronged effort of training and toolkit development,
PSIP contributed to improved methods and tools used by projects across ECP.
Examples are improvements in software builds and CI testing in the EXAALT materials modeling application,\footnote{\url{https://bssw.io/blog\_posts/adopting-continuous-integration-for-long-timescale-materials-simulation}}
mesh interface refactoring in Flash-X,\footnote{\url{https://bssw.io/blog\_posts/flash5-refactoring-and-psip}} and improvements by the HDF5 team in revision control, documentation, and coding standards.\footnote{\scriptsize \url{https://bssw.io/blog\_posts/recent-successes-with-psip-on-hdf5}} 
  
{\bf BSSw.io.} We established the Better Scientific Software (BSSw) website (\url{https://bssw.io})~\cite{bssw.io-2023} 
as a community-curated hub for sharing information about scientific software development.  Since the site's launch in 2017, over 275 contributors\footnote{\url{https://bssw.io/items/contributors}} have provided content covering a wide range of topics relating to scientific software planning, development, performance, reliability, collaboration, and skills, including original material created specifically for the site, curated links to external resources, and information on events relevant to the scientific software community.\footnote{Annual highlights showcase BSSw.io content, see \url{https://bssw.io/blog\_posts/better-scientific-software-2022-highlights} and counterparts for prior years.}

{\bf Collaboration via {\em teams of teams}.} 
Recognizing the importance of multiteam collaboration to address the challenges of next-generation computational science, we have advanced the concepts of {\em teams of teams}~\cite{TeamOfTeams} to strengthen community partnerships and scale productivity and innovation~\cite{10.1007/978-3-030-22338-0_39}. 
This approach has been particularly useful in fostering connections among the many multi-institutional teams that contribute to the scientific software ecosystem\footnote{\url{https://bssw.io/blog_posts/distributed-interconnected-teams-through-the-lens-of-team-of-teams-principles}} through SDKs and E4S, as well as providing a {\em team of teams} lens when considering work across ECP as a whole.
For example, the {\em team of teams} principles facilitated contributions of the HDF5 team to E4S and the Data and Viz SDK,\footnote{\url{https://ecp-data-viz-sdk.readthedocs.io/en/stable}} enabling advances in HDF5 functionality to readily support a variety of ECP applications, including modeling earthquakes, electronic structures, subsurface flow, reacting flow, stellar explosions, wind plants, and cosmology.
The opportunity to work closely with teams of teams in ECP also enabled advances in research on software repository mining and multi-team systems.\footnote{\url{https://www.osti.gov/biblio/1881691}} 

\section{Disseminating knowledge}
A central IDEAS goal is to establish a community-driven cycle in which widespread awareness of the importance of software quality and related issues in turn promotes sharing, discussion, and refinement of practices and resources for producing better scientific software. Key activities---all targeting the broad computational science community---include producing several webinar and panel discussion series, developing and delivering training, and providing recognition to leaders and advocates of high-quality scientific software.

In addition, we create opportunities for  informal 
cross-institutional dialogue and mechanisms to 
build community by sharing lessons learned.\footnote{\url{https://bssw.io/blog\_posts/think-locally-act-globally-outreach-for-better-scientific-software}} 
Few opportunities exist to discuss and learn about software engineering in 
science. A key part of the IDEAS outreach strategy is to enable the requisite peer knowledge sharing through workshops, focused sessions within larger conferences, and wherever possibilities present themselves throughout the scientific computing community.
These opportunities for peer knowledge sharing are helping advance research software engineering and {\em research software science}.\footnote{{\em Research software science} promotes the use of scientific methodologies to explore and establish broadly applicable knowledge related to research software engineering, {\doi{10.1109/MCSE.2023.3260475}}.} 
A few of our activities are described below.

{\bf Best Practices for HPC Software Developers (HPC-BP) Webinar Series.} Launched in 2016 in partnership with DOE advanced computing facilities (ALCF,\footnote{\url{https://www.alcf.anl.gov}} NERSC,\footnote{\url{https://www.nersc.gov}} and OLCF\footnote{\url{https://www.olcf.ornl.gov}}), the HPC-BP webinar series~\cite{jocse-10-1-19}
\footnote{\scriptsize \url{https://ideas-productivity.org/events/hpc-best-practices-webinars}} 
\footnote{\url{https://bssw.io/blog\_posts/best-practices-for-hpc-software-developers-the-first-five-years-of-the-webinar-series}}
was an immediate success, indicating an appetite in the HPC community to learn more about software engineering for science. The series
provides a monthly venue for topics related to scientific software development from practitioners throughout the international community. At the time of this writing, 71 webinars have been organized under ECP's auspices, presented by 87 speakers, with more than 12,000 registrations (about 37\% affiliated with ECP), a 48\% attendance rate on average, and about 2,300 unique attendees (considering attendees' email addresses). In the audience, 44\% were in the ``gov'' domain,
23\% in ``edu,'' 9\% in ``com,'' and the remaining attendees from other countries, gmail addresses, and so on.
Webinar recordings, slides, and questions/answers are posted online, enabling many more people beyond attendees of the "live" sessions to learn over time. 

{\bf Tutorials on scientific software practices.} Members of the IDEAS team regularly provide training at a variety of venues on topics of software testing, verification, revision control, refactoring, reproducibility, agile processes, and more.
All tutorials are available on our tutorial-specific website,\footnote{\url{https://bssw-tutorial.github.io}} including presentations and, when available, recordings.
Many, though not all, topics covered in the tutorial have a large body of knowledge as applied to commercial software, but very little for software used in science and research. 
One of the unique aspects of the scientific research community is the amount of education and experience a scientific software developer needs in a problem domain.  
Another difference is the nature of scientific discovery: software requirements emerge regularly and frequently, such that long-term detailed plans are always subject to change.
Tutorial modules either address topics unique to scientific software or are based on best practices in the broader software engineering community, tailored and adapted to the needs of the HPC scientific software community.

{\bf Better Scientific Software (BSSw) Fellowship Program.} To give recognition and funding to leaders and advocates of high-quality scientific software, we launched the BSSw Fellowship Program\footnote{\url{https://bssw.io/fellowship}} in 2018 with DOE support, with NSF joining sponsorship in 2021.
The main goal is to foster and promote practices, processes, and tools to improve developer productivity and software sustainability of scientific codes, while also focusing community attention on RSE contributions and providing a larger stage to advance causes related to high-quality software~\cite{GodoyEtAl-CiSE2023}. BSSw Fellows are selected annually based on an application process that includes the proposal of a funded activity (\$25k) that promotes better scientific software.  
A total of 27 BSSw Fellows in the 2018--2023 cohorts have developed training materials (including presentations in the {\em HPC-BP} webinar series and articles on BSSw.io) on topics such as code reviews, software testing, planning, design, and team collaboration; 27 BSSw Honorable Mentions have received recognition through community engagement.\footnote{\url{https://bssw.io/pages/meet-our-fellows}}

{\bf BoFs, workshops, and 
community events.} To foster discussion of issues in scientific software development, we (together with like-minded community members) typically organize sessions at scientific meetings and invite a
broad selection of speakers.\footnote{\url{https://ideas-productivity.org/events}}
We advertise these events widely and create archives to capture the events for future reference (e.g.,  BoF\footnote{Birds of a Feather  sessions provide a dynamic, non-commercial venue for conference attendees to openly discuss current topics of interest.}
sessions on software engineering for computational science\footnote{\url{http://bit.ly/swe-cse-bof}} at the SC and ISC conference series). We also organize standalone workshops that provide the opportunity for more in-depth interactions, notably the Collegeville Workshop on Scientific Software.\footnote{\url{https://collegeville.github.io/Workshops}}
\footnote{\url{https://bssw.io/blog\_posts/cultural-approaches-to-improved-software-teams-a-report-from-day-3-of-the-2021-collegeville-workshop-on-scientific-software}}

{\bf ECP panel series on performance portability.} 
As ECP project teams have been working toward performance portability across emerging exascale architectures, we partnered with DOE computing facilities and the three focus areas of ECP (application development, software technology, and hardware and integration) to lead an online panel series considering common themes of algorithmic and data locality challenges~\cite{perfportpanel2021}.

{\bf Panel series on Strategies for Working Remotely.} In response to the COVID-19 pandemic and the need for many in our community to transition to unplanned remote work, in spring of 2020 we launched the panel series {\em Strategies for Working Remotely},\footnote{\scriptsize \url{https://www.exascaleproject.org/strategies-for-working-remotely}} 
which explored experiences transitioning from co-located and partially distributed teams to fully virtual teams and teams of teams. Panelists have discussed challenges, lessons learned, and unforeseen benefits, as well as opportunities to work toward sustainable hybrid approaches for distributed collaboration in HPC.\footnote{\url{https://ascr-discovery.org/2022/06/work-shift}}

{\bf HPC Workforce Development and Retention: Action Group, webinar series, website.}  We initiated a webinar series\footnote{ \url{https://www.exascaleproject.org/workforce-development-seminar-series}} 
as part of the ECP Broadening Participation Initiative's\footnote{\url{https://www.exascaleproject.org/hpc-workforce}}
strategy to expand the pipeline and workforce for DOE high-performance computing. Led by the multi-lab HPC Workforce Development and Retention Action Group, webinars have addressed topics such as ally skills, diversifying computing, mentoring, and normalizing inclusion by embracing difference. A corresponding 
HPC-Workforce website\footnote{ \url{https://hpc-workforce-development-and-retention.github.io/hpc-wdr}} serves as a repository for webinar recordings, provides announcements of computing workforce events, and houses a growing collection of best practices on HPC workforce issues, such as the {\em inclusive minute}.\footnote{ \url{https://hpc-workforce-development-and-retention.github.io/hpc-wdr/jekyll/update/2023/04/08/inclusive-minute.html}}

\section{Achievements and opportunities}

The IDEAS team has pursued a multifaceted strategy to advance
scientific productivity through better scientific software.
Curating best practices for software development and team
productivity has empowered teams to build new
practices into their workflows and increase cross-project
collaboration.  
Software communities have proven to be
a source of inspiration for building shared foundations for software ecosystems while respecting team autonomy. 
IDEAS outreach mechanisms have enabled innovators in scientific software practices
to share knowledge with the community.

Feedback from ECP teams and others has highlighted the positive impact of IDEAS work. 
Firsthand accounts underscore the project's role in enhancing software quality, promoting best practices, and expanding awareness of the importance of software development. 
Many community members express a desire for additional resources, including tutorials on CI and tooling, support for CI/CD hardware, tactical guides for achieving scalability and portability, and more focus on debugging and post-mortems. 

The IDEAS multipronged strategy (to advance scientific productivity through better scientific software), as illustrated in Figure~\ref{fig:ideas-bigpicture},
is one that could be replicated in virtual organizations or local and regional ecosystems.
Key to its success is seeking cross-institutional collaboration around shared software
ecosystem needs, while considering 
use cases or user stories as a means to identify and ensure those needs are addressed.

{\bf Diffusion of innovations.} The evolution of ECP from a set of hardware
designs and science goals to a fully realized software ecosystem
with demonstrated performance on novel hardware taught us some
important lessons.  Access to and understanding of disruptive changes
are, by default, unevenly distributed.  These changes include both new
high-performance hardware and innovative ideas in software development---including processes such as CI testing, tools such as high-performance data visualization, and emerging strategies such as
active learning with AI in-the-loop.
One of the major successes of the ECP was in canvassing
the space of innovations while quickly disseminating those
technologies in a way that promoted community adoption and
contributions.
That level of integration---in hardware, software, and process---does
not happen on its own.  
It takes a dedicated, visible effort to foster a community that prioritizes adopting a shared scientific software development culture and to continue to catalyze its expansion and work for change. 

As the complexity of individual projects has grown,
the IDEAS project has met these challenges with innovations
in software process sustainability 
and advancement.  We addressed the diversity of (at times)
conflicting priorities with the concepts of aligning narratives
and liaisons (from the team of teams literature), and we promoted consensus-forming mechanisms to capitalize on
champions and early adopters
\cite{10.1007/978-3-030-22338-0_39}.
By building a sense of community in scientific software, we have
been able to maintain focus on the  shared goal to advance the
scientific frontier.
As the software needs for future-generation computational science are ever increasing, 
sustained investment is needed for community work to advance scientific software quality, productivity, and stewardship.
{\bf IDEAS as a catalyst.} The IDEAS project has always been a catalyst: fostering, incubating, curating, and disseminating best practices for scientific software development and use.  The IDEAS project also has served as a focal point for resources on scientific software practices that all teams can leverage.  We have been successful in advancing the state of the art in scientific software development, and we have helped  build a community of practice that is now poised to continue this work.

{\bf Moving forward.} While the IDEAS project will retain its role as a catalyst in the scientific software community, we believe two additional elements are essential for continued qualitative growth.  The first is increasing focus on {\em research software science},
enabling the community to explore and improve scientific software development and use via a scientific approach.  As scientists, we can benefit from using our scientific culture to accelerate the pace of improvement in scientific software. 
For example, building on our training in hypothesis-driven science, while incorporating a richer understanding of the social aspects of software development and use,
enables a combined social and traditional hard-science approach to addressing pressing software challenges in team-based science.

The second element needed is changing the prevailing community attitude toward software productivity and sustainability.  At present, a common attitude is that these are "nice to have" but not "must-haves." 
A core group of innovators and early adopters believe that software productivity and sustainability are "must-haves"---and we are working, in collaboration with others, to change culture and business models so that this important work is funded directly.  
Transparency and reproducibility---which require 
improvements in software productivity and sustainability---are needed for open, reproducible computational science to thrive and address next-generation challenges, where integrated research infrastructure and interdisciplinary simulation, analysis, and AI play such critical roles. 

We want all research sponsors in science and engineering to value improvements in software productivity and sustainability---and to expect that our community will use some of our base project funding to achieve those improvements.  Since many productivity and sustainability improvement efforts require long time spans to fully realize the benefits, deeply ingraining this culture change throughout the community will require investment at all organizational levels and evaluation of project success over a longer time span than standard funding cycles. We look forward to a future where continual software improvement is expected throughout our community, thereby accelerating reproducible and sustainable science and engineering.

\section{Acknowledgments}

This research was supported by the Exascale Computing Project (17-SC-20-SC), a collaborative effort of the U.S. Department of Energy, Office of Science, and the National Nuclear Security Administration.
We sincerely thank all those in the ECP, DOE, and broader communities of computational science and research software engineering who are  fostering communities of practice to help advance software quality in the pursuit of science.  We especially thank people who have contributed to the work discussed in this paper---by using and contributing to IDEAS resources, participating in community events, and providing motivation and feedback---so that together we are creating and diffusing innovation in scientific software practices to provide a firm foundation for robust and trustworthy next-generation computational science.  

\newcommand{\SIAMCSE}{SIAM Conference on Computational Science and Engineering,
  March 14-18, 2015, Salt Lake City, Utah}\newcommand{\SIAMGS}{SIAM Conference
  on Computational Issues in the Geosciences, June 29-July 2, 2015, Stanford,
  CA}\newcommand{\AGUFALLf}{AGU Fall Meeting, December 14--18, 2015, San
  Francisco, CA}\newcommand{\ESSPIf}{DOE Environmental Systems Science Annual
  Principal Investigators Meeting, April 28-29, 2015, Potomac,
  MD}\newcommand{\ESSPIs}{DOE Environmental Systems Science Annual Principal
  Investigators Meeting, April 26-27, 2016, Potomac, MD}\newcommand{\CMWR}{XXI
  International Conference Computational Methods in Water Resources, CMWR 2016,
  Toronto, ON, Canada.}\newcommand{\ESSWorkshop}{Environmental System Science
  {(ESS)} Workshop on Model-Data Integration, April 30 - May 1, 2015, Potomac,
  MD}

\clearpage
\bigskip
\bigskip
\section{Author bios}
\begin{IEEEbiography}{Lois Curfman McInnes}
is a senior computational scientist at Argonne National Laboratory. Her work focuses on scalable numerical libraries and community collaboration toward productive and sustainable scientific software ecosystems.  
\end{IEEEbiography}

\begin{IEEEbiography}{Michael A.~Heroux} is a senior scientist at Sandia National Laboratories and scientist in residence at St. John’s University, MN. His research interests include all human and technical aspects of scalable scientific and engineering software for new and emerging parallel computing architectures.\end{IEEEbiography}

\begin{IEEEbiography}{David E.~Bernholdt} is a distinguished R\&D staff member at Oak Ridge National Laboratory.  His research interests focus on the development of scientific software for high-performance computers, including developer productivity, and software quality and sustainability.
\end{IEEEbiography}

\begin{IEEEbiography}
{Anshu Dubey} is a senior computational scientist 
at Argonne National Laboratory and a senior scientist at the University of Chicago.
Her work focuses on software engineering for research, including the design, architecture, and sustainability of multiphysics applications.
\end{IEEEbiography}

\begin{IEEEbiography}
{Elsa Gonsiorowski} is an HPC support specialist at Lawrence Livermore National Laboratory and coordinator of the Better Scientific Software Fellowship program.
\end{IEEEbiography}

\begin{IEEEbiography}{Rinku Gupta}{\,}is a principal research software specialist at Argonne National Laboratory. Her work focuses on HPC software sustainability and productivity, fault tolerance, resiliency, and programming models.
\end{IEEEbiography}

\begin{IEEEbiography}
{Osni Marques} is a staff scientist at the Applied Mathematics and Computational Research Division, Lawrence Berkeley National Laboratory. His research interests include numerical linear algebra and high-performance computing.
\end{IEEEbiography}

\begin{IEEEbiography}
{David Moulton} is a senior scientist at Los Alamos National Laboratory.  His research focuses on the application of best practices software development methodologies and teams of teams approaches to the development of sustainable open-source software ecosystems.
\end{IEEEbiography}

\begin{IEEEbiography}{Hai Ah Nam} is a senior staff member at the National Energy Science Research Center at Lawrence Berkeley National Laboratory. She leads the NERSC-10 project to deliver the next-generation supercomputer for the DOE Office of Science research community.
\end{IEEEbiography}

\begin{IEEEbiography}
{Boyana Norris} collaborated on the IDEAS project while an associate professor in the Department of Computer and Information Science at the University of Oregon, focusing on enabling technologies for high-performance simulations. She recently transitioned to Rain Neuromorphics as a senior staff engineer, working on 
optimizing compilers for energy-efficient AI accelerators.
\end{IEEEbiography}

\begin{IEEEbiography}{Elaine M. Raybourn} is a social scientist at Sandia National Laboratories. Her research focuses on applying team science and diffusion of innovations to complex sociotechnical systems of teams of teams, HPC ethics, and incentivizing improvements in software quality and productivity. 
\end{IEEEbiography} 

\begin{IEEEbiography}
{Jim Willenbring} is a senior member of R\&D Technical Staff at Sandia National Laboratories. He is active in the research of software sustainability and the application of software engineering methodologies for high-performance computational science.
\end{IEEEbiography}


\begin{IEEEbiography}
{Ann Almgren} is a senior scientist at Lawrence Berkeley National Laboratory. Her research focuses on computational algorithms for solving partial differential equations in a variety of application areas.
\end{IEEEbiography}

\begin{IEEEbiography}{Roscoe A.~Bartlett} is a computational researcher and engineer at Sandia National Laboratories with 25+ years of experience in the R\&D of numerical methods and software for computational science and engineering (CSE).  His current focus is CSE software engineering challenges.
\end{IEEEbiography}

\begin{IEEEbiography}
{Kita Cranfill} is a software engineer for the Software Services Development Group at Oak Ridge National Laboratory. She designs and implements a variety of DOE software tools, including the HPC Workforce Development and Retention website.  
\end{IEEEbiography}

\begin{IEEEbiography}
{Stephen Fickas} is a professor of Computer and Information Science at the University of Oregon and the founder of Hop Skip Technologies, Inc.  He has a strong commitment to interdisciplinary research and education, including the quality of code, tests, and documentation. 
\end{IEEEbiography}

\begin{IEEEbiography}
{Don Frederick} is an HPC support specialist at Lawrence Livermore National Laboratory.  In the IDEAS project he liaises with Software Carpentry.
\end{IEEEbiography}

\begin{IEEEbiography}{William F. Godoy} is a senior computer scientist at Oak Ridge National Laboratory. His interests are in HPC scientific software infrastructure, programming models, and performance. 
\end{IEEEbiography}

\begin{IEEEbiography}{Patricia A.~Grubel} 
is a computer scientist in the Applied Computer Science group at Los Alamos National Laboratory. Her work includes the automation of scientific
workflows for reproducibility and provenance, containerization of applications, software engineering for sustainability, and profiling and optimization of scientific applications on new architectures.
\end{IEEEbiography}

\begin{IEEEbiography}
{Rebecca Hartman-Baker} leads the User Engagement Group at the National Energy Research Scientific Computing Center at Lawrence Berkeley National Laboratory. She is responsible for NERSC’s engagement with the user community to increase user productivity via advocacy, support, training, and the provisioning of usable computing environments.
\end{IEEEbiography}

\begin{IEEEbiography}
{Axel Huebl} is a computational physicist at Lawrence Berkeley National Laboratory. He works at the interfaces of HPC, laser-plasma physics, and advanced particle accelerator research developing accelerators, including development of WarpX (2022 Gordon Bell Winner) and drives open standards in his community.
\end{IEEEbiography}

\begin{IEEEbiography}
{Rose Lynch} is an administrative assistant at Argonne National Laboratory. She helps the IDEAS team stay organized and communicate clearly.  
\end{IEEEbiography}

\begin{IEEEbiography}{Addi Malviya-Thakur} serves as the Software Engineering Group leader in the Computer Science and Mathematics Division at Oak Ridge National Laboratory. Her interests include interconnected science and federated systems, research software engineering,  workflows, software frameworks, and ecosystems for science. 
\end{IEEEbiography}

\begin{IEEEbiography}{Reed Milewicz} is a computer scientist and software engineering researcher in the Department of Software Engineering and Research at Sandia National Laboratories, where he is the research lead. His research focuses on developing better practices, processes, and tools to improve software development for science.
\end{IEEEbiography}

\begin{IEEEbiography}{Mark C.~Miller} has 30+ years experience contributing to scientific data analysis and modeling technologies such as Silo, ASCI-DMF, HDF5, ITAPS, and VisIt. His interests include data models, interoperability, scalable I/O, and software quality engineering.
\end{IEEEbiography}

\begin{IEEEbiography}{Miranda R. Mundt} is a research software engineer at Sandia National Laboratories. She supports scientific software development and conducts fundamental software engineering research in software sustainability.
\end{IEEEbiography}

\begin{IEEEbiography}{Erik Palmer} is a software integration engineer at the National Energy Science Research Center at Lawrence Berkeley National Laboratory. He works in the User Engagement Group, consulting with users on HPC issues, building software, and coordinating user-facing CI efforts.  
\end{IEEEbiography}

\begin{IEEEbiography}{Suzanne Parete-Koon} is an HPC engineer at Oak Ridge National Laboratory and leads their user training effort for the Frontier exascale supercomputer.  She also leads the HPC Workforce Development and Retention Action Group. 
\end{IEEEbiography}

\begin{IEEEbiography}{Megan Phinney} 
is a computer scientist in the High Performance Computing -- Environments group at Los Alamos National Laboratory, where she provides user support for the institutional HPC systems.  Her research interests include secure user space container environments and performance of file systems.
\end{IEEEbiography}

\begin{IEEEbiography}{Katherine Riley} is the director of science for the Argonne Leadership Computing Facility. She leads a team of experts in computational science, performance engineering, visualizations, and data sciences. 
\end{IEEEbiography}

\begin{IEEEbiography}
{David M. Rogers} is a computational scientist in the National Center
for Computational Sciences Division, Oak Ridge National Laboratory.
His research interests include computation, mathematics, and theory
enabling HPC methods for electron- to fluid-scale modeling.
\end{IEEEbiography}

\begin{IEEEbiography}{Benjamin Sims} is a sociologist at Los Alamos National Laboratory. His research focuses on how communities of scientists, engineers, software developers, and others come together to build and maintain infrastructures and scientific knowledge.
\end{IEEEbiography}

\begin{IEEEbiography}{Deborah Stevens} is a computer scientist at Argonne National Laboratory with a background in applied mathematics and numerical methods. Her recent work focuses on promoting software sustainability. \end{IEEEbiography}

\begin{IEEEbiography}{Gregory R.~Watson}
is  leader of the Application Engineering
Group in the Computer Science and Mathematics Division at Oak Ridge National Laboratory.
His research interests include programming tools and development
environments for high-performance and scientific computing, software
engineering practices, reproducibility, and training for scientific computing.
\end{IEEEbiography}

\end{document}